\documentclass[utf8]{frontiersSCNS} 

\usepackage{url,hyperref,lineno,microtype,subcaption}
\usepackage{comment}
\usepackage[onehalfspacing]{setspace}
\usepackage{caption}
\usepackage{subcaption}
\usepackage{float}
\usepackage{graphicx}
\usepackage{fancyhdr}
\usepackage{pgfplots}

\def\keyFont{\fontsize{8}{11}\helveticabold }
\def\firstAuthorLast{Fosque {et~al.}} 
\def\Authors{L. J. Fosque\,$^{1}$, A. Alipour\,$^{2}$, M. Zare\,$^{3}$, R. V. Williams-Garc{\'i}a\,$^{4}$, J. M. Beggs\,$^{1}$ and G. Ortiz\,$^{1,*}$}

\def\Address{$^{1}$ Department of Physics, Indiana University, Bloomington, IN, 47405, USA \\
$^{2}$  Department of Psychological \& Brain Sciences, Indiana University, Bloomington, IN, 47405, USA \\
$^{3}$ AI Practice, CGI, Montreal, QC H3G 1T4, Canada \\
$^{4}$ Institut Denis Poisson (UMR7013), Parc de Grandmont, F-37200 Tours, France} 

\def\corrAuthor{Gerardo Ortiz}

\def\corrEmail{ortizg@iu.edu}

\begin{document}

\onecolumn
\firstpage{1}

\title{Quasicriticality explains variability of human neural dynamics across life span} 

\author[\firstAuthorLast ]{\Authors} 
\address{}

\correspondance{}
\extraAuth{}

\maketitle

\begin{abstract}

Ageing impacts the brain's structural and functional organization and over time leads to various disorders, such as Alzheimer's disease and cognitive impairment. The process also impacts sensory function, bringing about a general slowing in various perceptual and cognitive functions. Here, we analyze the Cambridge Centre for Ageing and Neuroscience (Cam-CAN) resting-state magnetoencephalography (MEG) dataset --the largest ageing cohort available-- in light of the quasicriticality framework, a novel organizing principle for brain functionality which relates information processing and scaling properties of brain activity to brain connectivity and stimulus. Examination of the data using this framework reveals interesting correlations with age and gender of test subjects. Using simulated data as verification, our results suggest a link between changes to brain connectivity due to ageing, and increased vulnerability to distraction from irrelevant information. Our findings suggest a platform to develop biomarkers of neurological health. 



\tiny
 \keyFont{ \section{Keywords:} Magnetoencephalography (MEG), Quasicriticality, Cam-CAN dataset, Healthy Human, Aging, Neuronal Avalanches} 
\end{abstract}

\section{Introduction}

It is extremely useful in medicine to have biomarkers that will diagnose or predict a patient’s health (\cite{Montez2009, Zijlmans2012, Mena2016, Bruining2020}). In neuroscience, the rapid advancement of data collection and analysis has generated many new candidate biomarkers, like in seizure prediction, for example (\cite{Scheffer2009, Kuhlmann2018}). However, it is sometimes difficult to tell if a new biomarker underlies the cause of a condition or merely has a spurious correlation with it (\cite{Granger1974, Mormann2007, Calude2017}). A strategy to avoid this pitfall would be to focus on observables that are tied to homeostatic functions that are known to establish conditions for health.

A growing body of work suggests that the brain homeostatically regulates itself to operate near a critical point where information processing is optimal (\cite{Ma2019, meisel2017decline, Beggs2022book}). At this critical point, incoming activity is neither amplified (supercritical) nor damped (subcritical), but approximately preserved as it passes through neural networks (\cite{Beggs2008}). Departures from the critical point have been associated with conditions of poor neurological health like epilepsy (\cite{Meisel2012}), Alzheimer’s disease (\cite{Montez2009})  and depression (\cite{Gartner2017}; for a review, see \cite{Zimmern2020}). For example, when subjects are deprived of sleep, their brain signals as assessed by electroencephalography (EEG) move from near the critical point towards being supercritical, where there is an increased likelihood of seizures. After restorative sleep, subjects move back toward being subcritical where seizures are less likely (\cite{Meisel2013}). Homeostasis of criticality has been observed in animal experiments too, where prolonged eye suture causes visual cortex to become subcritical. After several days under this condition, the cortex returns toward the critical point (\cite{Ma2019}). These and other studies (\cite{Tetzlaff2010,shew2015turtle, fontenele2019criticality}) indicate that there are mechanisms to maintain the brain near the critical point, even in the face of strong perturbations. 

The brain, however, is never exactly critical and it is natural to ask if there is an organizing principle behind the neocortex's functional behavior. The quasicriticality framework (\cite{williams-garcia2014}) has been advanced as such a principle. A new way to plot the brain’s proximity to the critical point, rooted in this idea of quasicriticality, has revealed that the effective critical exponents (explained more below) are confined to lie near a scaling line (\cite{friedman2012universal}). For example, as a rat explores, grooms or sleeps over several hours, its effective critical exponents may change, but they consistently lie near this scaling line, moving along it over time (\cite{fontenele2019criticality}). In another experiment that illustrates this, as rats learned a lever pressing task over several weeks their effective critical exponents moved along the scaling line but never strayed far from it (\cite{Ma2020}). Similar findings have now been reported by several labs, including our own (\cite{shew2015turtle,fosque2021evidence}). This makes it reasonable to ask if a new biomarker for neurological health could be based on the brain’s tendency to operate within the quasicritical region. Based on this work, we hypothesize that the position, and its change,  along this scaling line could serve as a biomarker for neurological health. 

To pursue this, we examined a large set of magneto-encephalography (MEG) data collected from 604 healthy human subjects. We extracted the effective critical exponents from each patient and plotted them along the scaling line. As a first step toward testing this idea, we asked if the position on the scaling line contained information about the patients’ age, gender and sensitivity to inputs. We found that statistically significant relationships did exist, suggesting that this approach will be fruitful for other health-related information. 

To quantitatively interpret these results, we used a previously-published computational model of brain dynamics, the so-called {\it cortical branching model} (CBM) (\cite{williams-garcia2014}). When we supplied this model with activity levels and connectivity patterns that approximated those from the patient population, it produced outputs that were consistent with the results from the human data. This model gives us an intuitive understanding of why the effective critical exponents move along the scaling line and how they might be related to departures from criticality. 

The remainder of this paper is organized as follows. In the next section we will briefly explain some background information related to quasicritical behavior, like effective critical exponents, the scaling relation and scaling line, as some of these concepts are imported from physics and are not widely known in neuroscience. After that, we will describe the methods of data collection and analysis. This will be followed by the results section, which describes the data analysis and computational modeling. Lastly, we will discuss possible limitations and future applications of this approach.  

\section{Background: The organizing principle of quasicriticality}

Broadly speaking, activity in neural networks propagates in successive stages, spreading from one set of active neurons to another, and resulting in spatio-temporal patterns of activation known as neuronal avalanches. This propagation can be quantified using the branching ratio $\sigma$, which gives the average number of neurons activated by a single active neuron. If we consider a network with infinitesimal inputs, then the network is supercritical when $\sigma >1$, and incoming signals are successively amplified. After several stages, it is difficult to discern from the output which neurons were active at the input because the network becomes nearly saturated with activity. When $\sigma <1$, the network is subcritical and incoming signals are damped. After several stages, it is also difficult to discern which neurons were active at the input, now because there is almost no activity at the output. When $\sigma \approx 1$, the network nears a critical point where levels of activity are roughly maintained. In this condition, it is easiest to use the output to reconstruct the input; mutual information between inputs and outputs is maximized for critical systems (\cite{shew2011information}). For similar reasons, a network’s dynamical response to inputs, i.e., its susceptibility, is also maximized near the critical point and diverges at the critical point $\sigma=1$. It has become clear that this picture seems consistent with experimental evidence (\cite{shew2011information,Wilting2019}). This type of evidence has been obtained from \textit{in vitro} culture preparations (\cite{beggs2003neuronal}) as well as \textit{in vivo} recordings from fish (\cite{Ponce2018}), rodents (\cite{fontenele2019criticality}), primates (\cite{Petermann2009}),  and humans (\cite{Tagliazucchi2012, shriki2013neuronal}). 

\begin{figure}[h!]
    \centering   \includegraphics[scale=0.5]{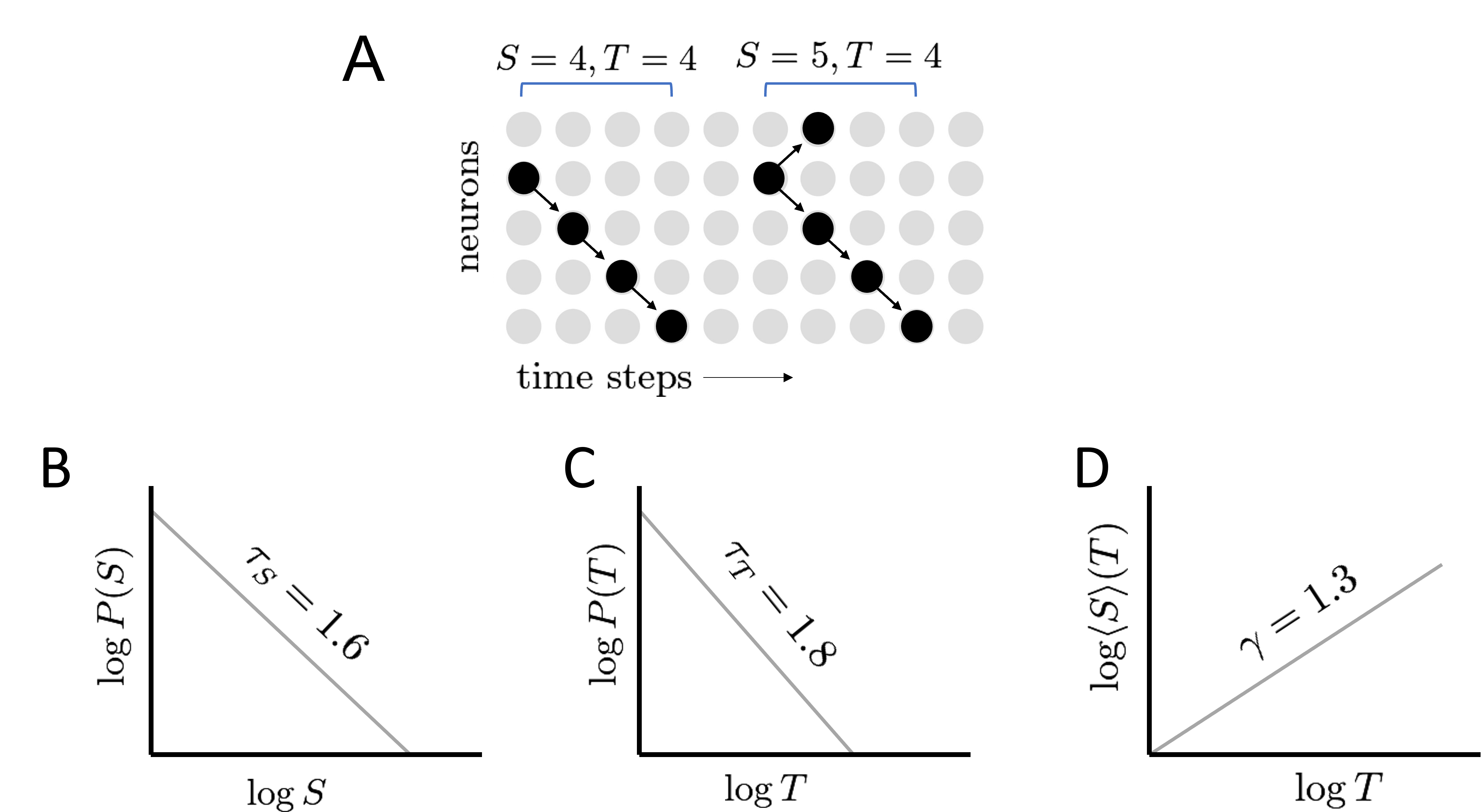}
    \caption{Neuronal avalanches and their distributions. \textbf{A:} Schematic network of 5 neurons; inactive neurons are represented by gray dots and active neurons by black dots; and propagation of activity is represented by arrows. Two avalanches are shown: the first with size 4 and duration 4, and the second with size 5 and duration 4. \textbf{B:} In a critical network, the distribution of avalanche sizes $S$, follows a power law, which appears linear when plotted with logarithmic scales. The slope of this line gives the exponent $\tau_S= 1.6$. \textbf{C:} The distribution of avalanche durations, $T$, follows another power law with exponent $\tau_T = 1.8$. \textbf{D:} The average avalanche size for a given duration, $\langle S \rangle(T)$, follows a power law with exponent $\gamma=4/3$. The appearance of multiple power laws and the relationship between their exponents, $(\tau_T -1)/(\tau_S -1)=\gamma$, indicate that the network may be operating near a critical point.}
    \label{fig:Avalanches}
\end{figure}

Other signatures of criticality include the scale-free distributions of measured quantities, for example, the size and duration of neuronal avalanches.  Close to criticality, avalanche size and duration distributions nearly follow power laws, showing approximate scale invariance (Fig. \ref{fig:Avalanches}) \cite{BookNO}. The fact that these distributions are nearly scale-free indicates the absence of a dominant length or time scale, which would be small in the case of subcritical networks and large in the case of supercritical networks. Relations between the characteristic exponents of these power laws help to confirm critical behavior. For instance, the network produces cascades, or avalanches, of activity whose sizes $S$ and durations $T$ follow power law distributions, $P(S)\propto S^{-\tau_S}$ and $P(T)\propto T^{-\tau_T}$, respectively. When represented in a log-log plot, power law distributions appear as straight lines and the slopes of these distributions are used to estimate critical exponents. When the average avalanche size $\langle S\rangle$ is plotted against avalanche duration, this also produces a straight line in a log-log plot; the exponent for this is $\gamma$, and the distribution $\langle S\rangle\propto T^{\gamma}$.
At criticality, the avalanche size exponent $\tau_S$ and avalanche duration exponent $\tau_T$ are related by the exact scaling relation (\cite{BookNon-Eq, sethna2001crackling,friedman2012universal}), 
\begin{eqnarray}
\gamma = \frac{\tau_T-1}{\tau_S-1} ,
\label{gammascaling}
\end{eqnarray}
where $\gamma$ corresponds to the characteristic exponent of average avalanche size for a given duration, $\langle S \rangle(T)$, another scale-free quantity. This scaling relation will not be satisfied away from the critical point, and it will only be {\it approximately} satisfied near it. It can therefore be used as a way to assess proximity to the critical point. The interesting questions are (1) How close is close? (2) How close can a neural network be to the critical point? and, most importantly,  (3) Is there an organizing principle for neural dynamics?

\begin{figure}[htb]
    \centering
    \includegraphics[scale=0.38]{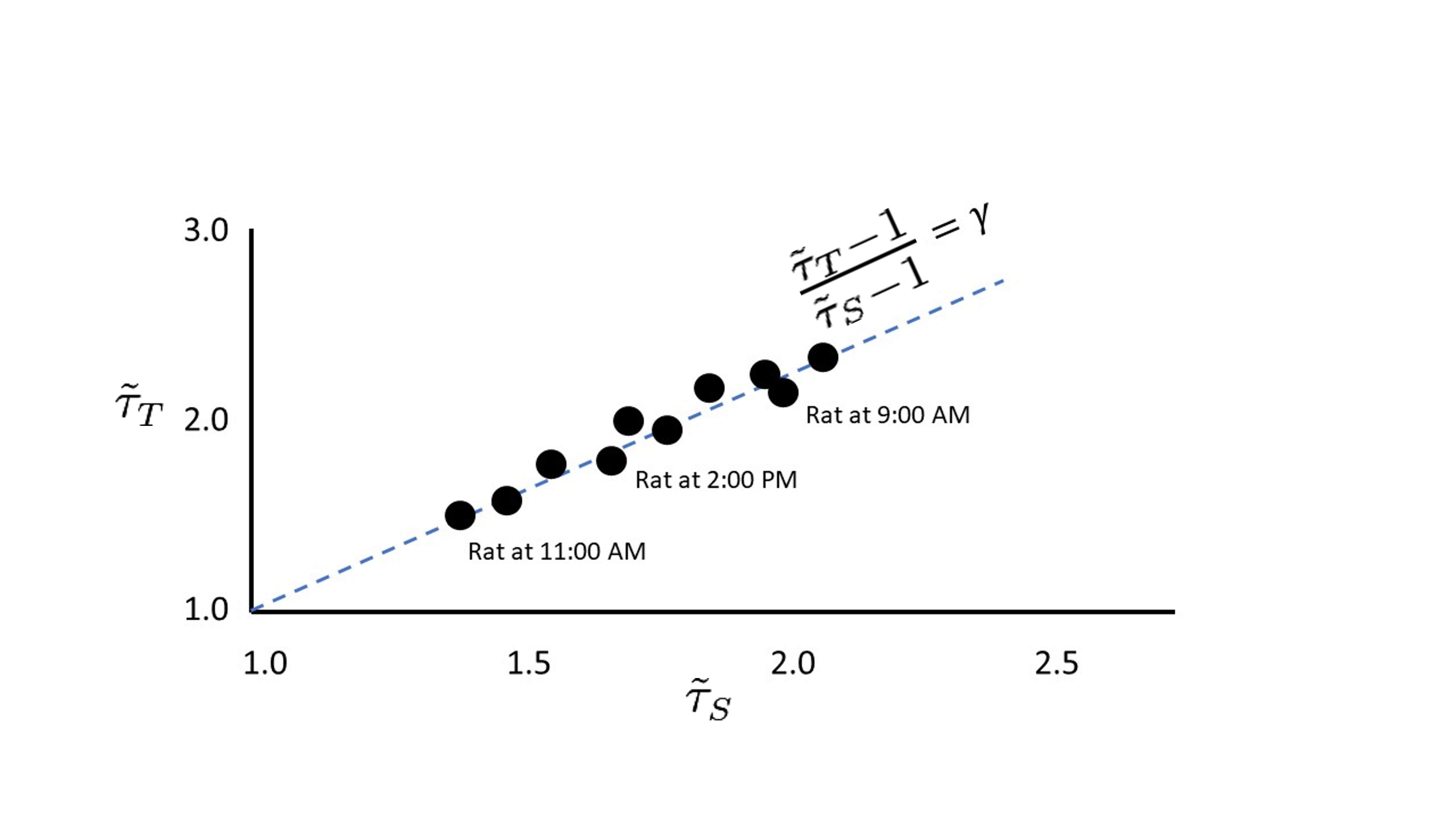}
    \caption{Empirical data stay near the $\gamma$-scaling line. Depiction of avalanche duration exponents, $\tilde \tau_T$, plotted against avalanche size exponents, $\tilde \tau_S$ for data collected from the same rat at different times (similar to data presented by \cite{fontenele2019criticality}). Note that in all cases, exponents lie close to the dashed $\gamma$-scaling line given by $\gamma = (\tilde \tau_T-1)/(\tilde \tau_S-1)$. Proximity to the $\gamma$-scaling line indicates closeness to the critical point.}
    \label{fig:Scalingline}
\end{figure}

The quasicriticality hypothesis states that living neural systems, being constantly bombarded by external input, will adapt to operate in the functional parameter region near the peak of maximum dynamical susceptibility of neural activity, \cite{williams-garcia2014}. This quasicritical region maximizes information propagation across the neural network. Moreover, the peak of maximum susceptibility defines a non-equilibrium Widom line (or surface) that depends on the level of external noise (or stimulus) as well as other functional parameters of the network. It is important to emphasize that the quasicritical hypothesis represents a universal organizing principle and not a particular non-equilibrium model of brain cortex dynamics (such as the CBM, \cite{williams-garcia2014}). As mentioned earlier, there is now evidence of homeostasis towards a {\it quasicritical} region, even after strong disruptions. In this sense, homeostasis of neuronal activity is analogous to homeostasis of blood pressure, heart rate, and body temperature. Because it is useful to track these vital signs over time, it would also make sense to track proximity to this quasicritical region over time. A new way to do this has emerged from the previously-mentioned exponent relation and the organizing principle of quasicriticality (\cite{williams-garcia2014, fontenele2019criticality, fosque2021evidence}). In this framework, although the network is not exactly at the critical point, it may still have approximate power-law distributions with {\it effective critical exponents}. 
When the network operates away from the critical point, the two sides of Eq. \eqref{gammascaling} will not {\it exactly} equal each other. The magnitude of this difference has been called the distance to criticality coefficient (DCC) \cite{Ma2019}.  At, or near, the Widom line this relation will be approximately satisfied \cite{williams-garcia2014,fosque2021evidence} providing evidence for quasicritical behavior. 
The effective avalanche duration exponent, $\tilde \tau_T$, and the effective avalanche size exponent, $\tilde \tau_S$ can be plotted as a point in the $(\tilde \tau_T,\tilde \tau_S)$ plane. Over time, depending on stimuli, noise, and intrinsic structural parameters of the neural network, it is found that while $\tilde\tau_T$ and $\tilde \tau_S$ may change, they typically remain close to the scaling line, the so-called $\gamma$-scaling line,  given by exact scaling relation. This is shown schematically in Fig. \ref{fig:Scalingline}. 
 
\begin{figure}[h!]
    \centering
    \includegraphics[scale=0.4]{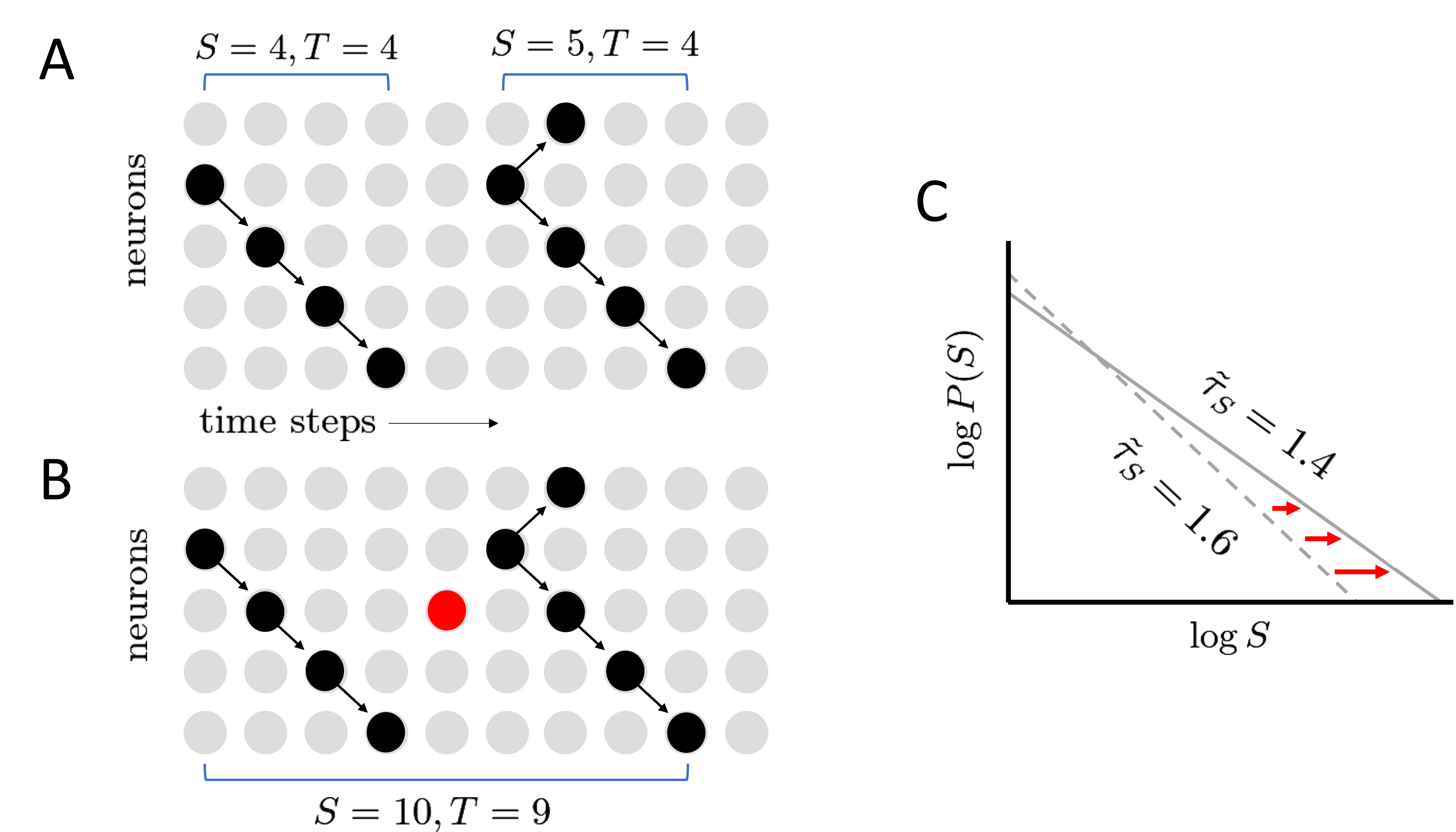}
    \caption{How spontaneous activity reduces exponent magnitude. \textbf{A:} Two avalanches as shown before. Note gap in activity separating them. \textbf{B:} Spontaneous activity turns on neurons that would otherwise have been inactive (red circle), filling the gap. The two previously distinct avalanches are now merged together into a larger and longer avalanche. \textbf{C:} The increase in large avalanches causes the tail of the distribution to move further to the right. This in turn causes the magnitude of the exponent $\tilde \tau_S$ to decrease (1.6 to 1.4). Figure illustrates schematically the results of quasicritical simulations (\cite{fosque2021evidence}). }
    \label{fig:Spontaneous}
\end{figure}

\begin{figure}[h!]
 \includegraphics[width=\linewidth]{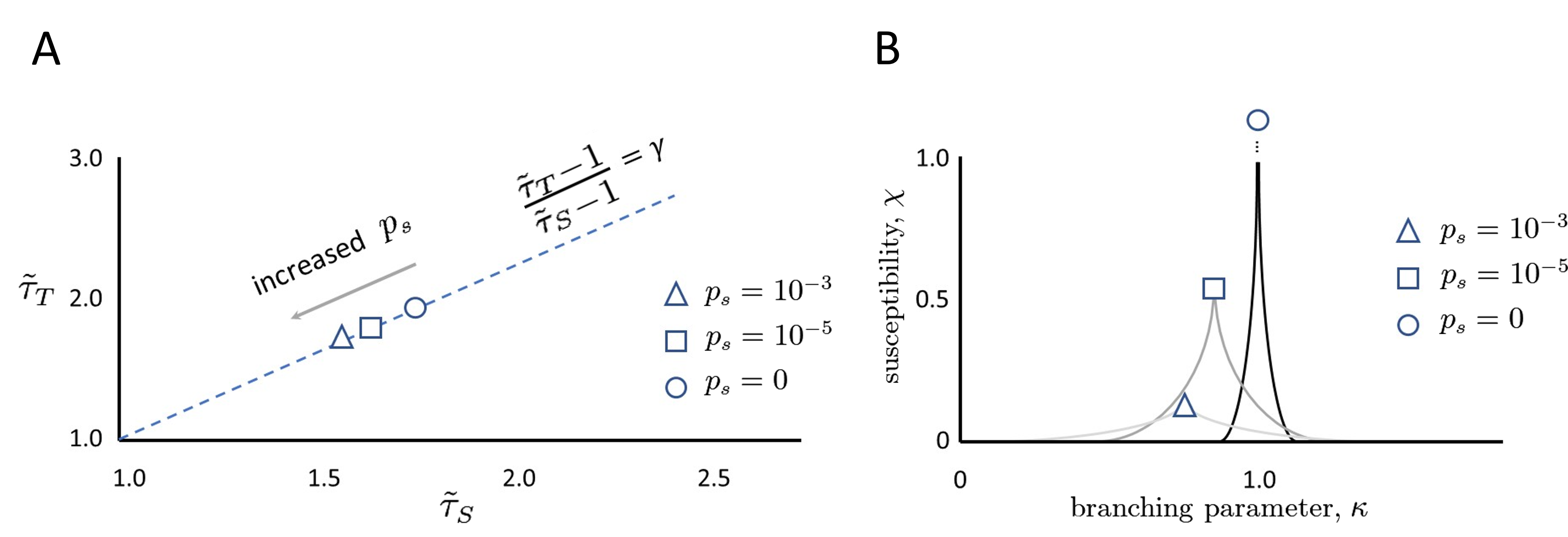}
    \caption{ \textbf{A:} Moving along the $\gamma$-scaling line. Avalanche duration exponents, $\tilde \tau_T$, plotted against avalanche size exponents, $\tilde \tau_S$ for different values of $p_s$. Note that in all cases, exponents lie close to the dashed $\gamma$-scaling line given by the equation $\gamma=(\tilde \tau_T -1)/(\tilde \tau_S -1)$. When there is no spontaneous activity, the magnitude of the exponents is largest (circle). As $p_s$ is increased, the magnitude of the exponents decreases (square, then triangle). Schematic plot here summarizes results reported by (\cite{shew2015turtle}; \cite{fontenele2019criticality}; \cite{fosque2021evidence}) and predicted by models of quasicriticality (\cite{williams-garcia2014}; \cite{fosque2021evidence}).  \textbf{B:} Susceptibility, $\chi$, is blunted by increases in $p_s$.  When $p_s=0$ and the branching parameter is exactly one, the susceptibility curve will diverge to infinity (circle). As $p_s$ is increased (square, triangle), the susceptibility declines. Note that the branching parameter at which these curves peak also declines. This figure schematically depicts the predictions of a quasicritical network model (\cite{williams-garcia2014}) that have recently been corroborated with data from spiking networks (\cite{fosque2021evidence}).}
    \label{fig:Movingline}
\end{figure}

What things can push a network away from the critical point? Randomness is ubiquitous in the nervous system, from synaptic transmission to thermal noise causing neurons to spontaneously fire (\cite{White2000}). When a perfectly critical network experiences noise, it moves away from the critical point in very specific ways. Intuitively, when the probability of spontaneous activity in neurons, $p_s$, is increased, formerly distinct avalanches are concatenated (Fig. \ref{fig:Spontaneous}) as shown in (\cite{williams-garcia2014}). This increases the number of large avalanches, which decreases the slope of the distribution, and thus reduces the effective exponent. This can account for movement along the $\gamma$-scaling line toward the origin, as shown in Fig. \ref{fig:Movingline}. This contrasts with what is expected to happen exactly at the critical point, where only one set of exponents should be found - a single point that does not move on the line - since that point (together with other exponents) represents a single universality class (\cite{BookNO}). Thus, moving exponents indicate that the network is not critical at all. But if the exponents move along the line, then the network is as close to critical as it could be - this is the origin of the term quasicriticality.

It is important to note that this phenomenon of avalanche concatenation will radically change the value of $\sigma$ as traditionally measured, due to absence of separation of timescales between driving and relaxation processes. Hence, we use the branching parameter $\kappa$, the largest eigenvalue of the connectivity matrix, as the control parameter that tunes the system close to the peak of susceptibility. Note that $\kappa$ can be considered to be the theoretical branching ratio, while $\sigma$ is the empirical one measured without distinguishing concatenated avalanches, and in the critical case when $p_s=0$ both will be unity. This parameter, and its relation to the branching ratio $\sigma$, is explained in Methods.

\begin{figure}[h!]
    \centering
    \includegraphics[scale=0.4]{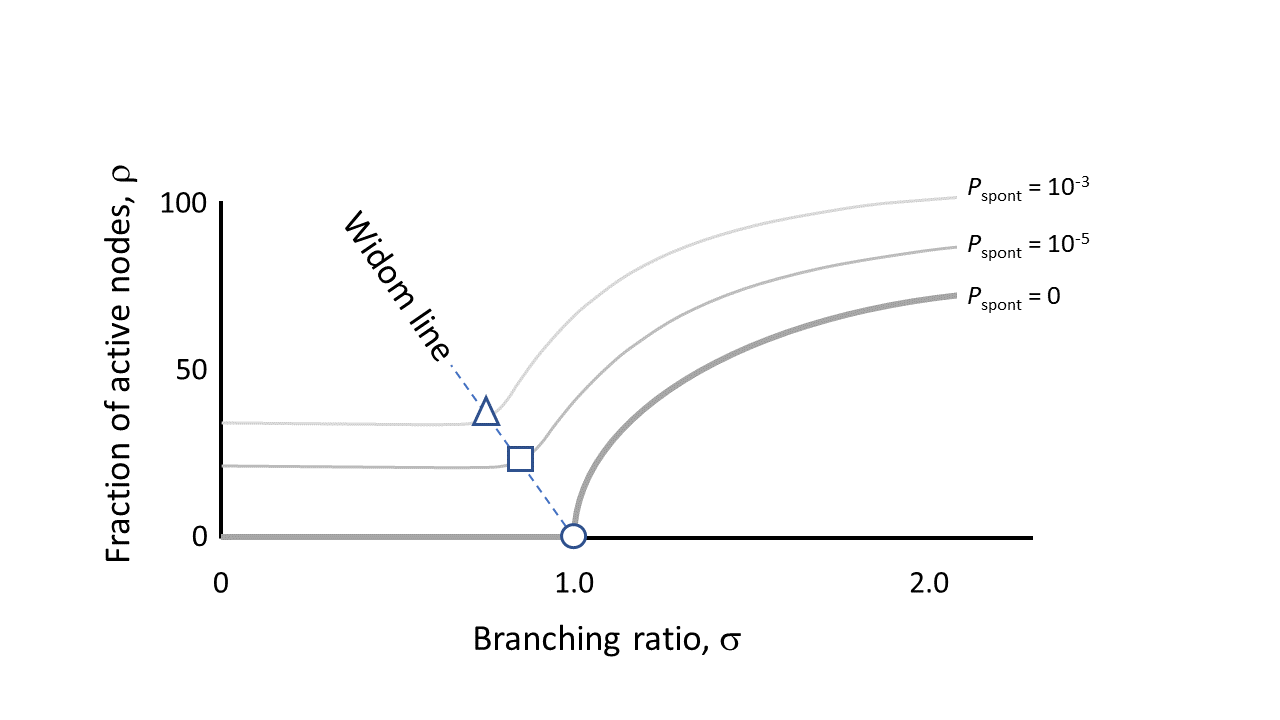}
    \caption{Phase diagram illustrating quasicriticality and the Widom line. When $p_s=0$, there is an inactive phase to the left of the critical point ($\kappa=1=\sigma$) and an active phase to the right of it. The critical point is given by the circle. The thick gray curve shows how the fraction of active neurons (y-axis) varies as the control parameter $\kappa$ is increased. As $p_s$ is increased (lighter gray curves), this curve is shifted vertically and to the left. The points at which susceptibility will be maximal for each value of $p_s$ are given by the square and the circle. While the network will not be critical at these points, it will be quasicritical (quasicritical region is enclosed with a parabolic black dashed line). This means that susceptibility will be maximal for that level of spontaneous activity. The dashed line joining these optimal points is called the Widom line. Note that the branching parameters at which maximum susceptibility occurs are now shifted to the left. We can also see that there are three main regions divided by the Widom line, the subcritical and supercritical ones. This figure schematically depicts predictions of a quasicritical network model (\cite{williams-garcia2014}) which have recently been corroborated with data from spiking networks (\cite{fosque2021evidence}).}
    \label{fig:PhaseDiagram}
\end{figure}

An increase in $p_s$ has other effects as well. As there is more random activity, mutual information between inputs and outputs will be reduced; this is like experiencing static in a phone call and being less able to hear the speaker. 
Moreover, because the network will be less responsive to slight changes in the inputs, the susceptibility of the network will also decrease with increased $p_s$ (Fig. \ref{fig:Movingline}).
To better illustrate, we draw a non-equilibrium phase diagram of the system (Fig. \ref{fig:PhaseDiagram}) using the branching parameter $\kappa$ as the control parameter. When $p_s=0$, the critical point occurs at $\kappa=1$ and there are two distinct phases: the inactive subcritical phase ($\kappa<1$) and active supercritical phase ($\kappa>1$). 
When $p_s$ is non-zero, however, the phase transition is replaced by a crossover region between less active and more active network states; there is always some network activity, even in the less active states. This crossover region is in fact the quasicritical region and contains the peak susceptibility for a given value of $p_s$. The values of $\kappa$ at which the susceptibility peaks for each value of $p_s$ are themselves connected by the (dynamical) Widom line.


With large changes in $p_s$, a network may need to ``move'' back to the quasicritical region by adjusting $\kappa$. As shown in \cite{fosque2021evidence}, there is a correspondence between this ``adjustment" and ``movement" along the $\gamma$-scaling line, due to the change in the values of effective exponents $\tilde \tau_T$ and $\tilde \tau_S$ as $p_s$ is changed. This is the underlying mechanism responsible for the movement of exponents along the $\gamma$-scaling line in Figs. \ref{fig:Scalingline} and \ref{fig:Movingline}. This picture of moving exponents fits well with recent reports that the exponents of a given animal can move along this exponent relation line over time \cite{fontenele2019criticality}  or after learning \cite{Ma2020}. Recent experiments with cortical slice cultures are also consistent with this framework \cite{fosque2021evidence}. Thus, although a quasicritical network is not at the critical point it is, in the sense described above, ``close enough'' to still enjoy improved information processing compared to networks that are arbitrarily away from the critical point \cite{helias2021brain}.
With this, we hypothesize that human MEG data will also fall along the $\gamma$-scaling line. Most importantly, the position along the $\gamma$-scaling line may reveal useful information about patient health and age. In this paper, we take the first steps toward exploring this idea. Can the position along the $\gamma$-scaling line be linked with basic patient information like age, gender, or sensitivity to new inputs?

\section{methods}
\subsection{Data and Participants}

We analyzed resting-state MEG data of 604 participants (304 male and 300 females) aged 18-88 years old provided in the Cam-CAN database. For each subject, temporal Signal Space Separation (tSSS, MaxFilter 2.2, Elekta Neuromag, Oy, Helsinki, Finland) was applied to remove noise from external source and from HPI coil. The sampling frequency of recorded data was 1 kHz with a high-pass filter of 0.03 Hz. Independent component analysis had been performed by Cam-CAN to exclude signal components associated with eye movements. The resting-state recordings were each at least 8 min and 40s in duration.  The MEG sensor array consisted of 306-channel Elekta Neuromag Vectorview (102 magnetometers and 204 planar gradiometers). We analyzed the magnetometer sensors only- although we have analyzed all directions, the magnetometer time-series provided better and more pronounced avalanche distributions.  This gave us as a result 102 channels of time-series activity for each participant. More details about the data acquisition pipeline can be found in \cite{taylor2017cambridge}. Computations were carried out using MATLAB (R2020a, The Mathworks Natick, MA), and the Python programming language (Python Software Foundation. Python Language Reference, version 3.8. available at \url{https://www.python.org/}), GNU Parallel \cite{Tange2011a}.

\subsection{Extracting Neuronal Avalanches from MEG sensor activity}

\begin{figure}[h!]
    \centering
    \includegraphics[width=0.8\linewidth]{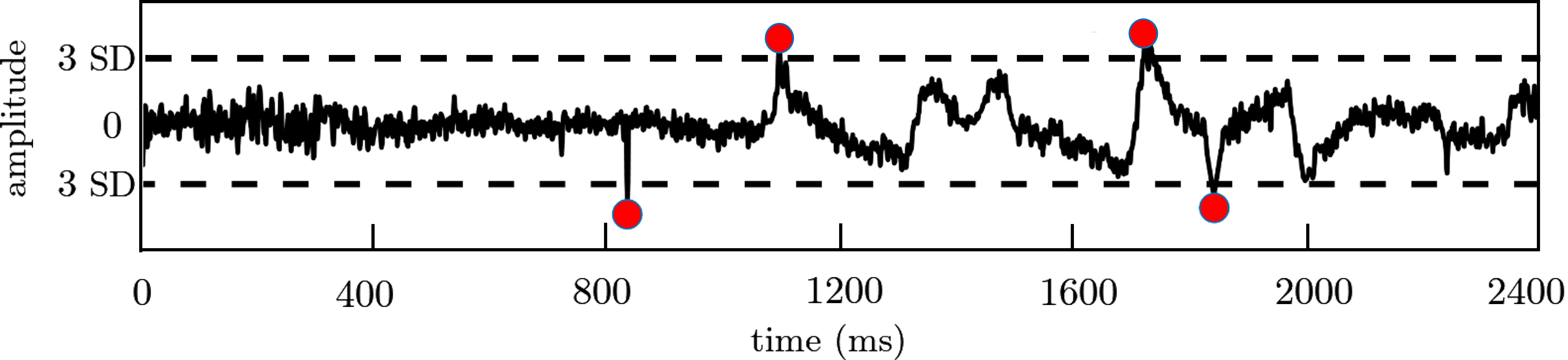}
    \caption{Thresholding of a time-series for one channel using 3 SD. As illustrated in the figure, this node/channel $z_i(t)$ was active, i.e., $z_i(t_n)=1$, in 4 occasions, $t=t_n$,  during this time interval.}
    \label{thresholding}
\end{figure}

In MEG data, each channel's activity is presented in the form of a continuous-time series. The amplitudes, positive or negative, of these time-series determine the level of neural activity in the corresponding channel. To determine the significant, active, events we set a threshold  separating relevant activity from background noise. In this way, we map to a discrete-time series. The method used amounts to:
\begin{itemize}
    \item Let $N=102$ be the total number of nodes/channels per subject, and $z_i$ the state of the node $i$, $i=1,\cdots,N$. A node can have only one of two possible states $z_i(t)\in \{0,1\}$ at a given time $t$.
    \item To determine the state of the node $z_i(t)$ from MEG data, we first 
    subtract the mean activity and divide it by the standard deviation (SD). We then threshold the resulting time series (using SD as shown in Fig. (\ref{thresholding})) so that events, positive or negative, that surpass the threshold level represent the active $z_i(t_n)=1$ state while the rest is considered inactive, i.e., $z_i(t)=0$, for times $t\neq t_n$. We adjust the threshold by following the same procedure as in  \cite{dehghani2012avalanche} and \cite{jannesari2020}. 
\end{itemize}

 We found that a threshold much lower than 3 SD led to the detection of many noisy events, and 3 SD showed to be optimal for most subjects. This discretization map is similar to the one used in previous works (\cite{shriki2013neuronal}, \cite{jannesari2020}). 

Once we established the activity map, we need to set relevant time scales in the time series in order to study the probability distributions of avalanches. This is known as the binning process. We binned datasets into different inter-event separation times $\Delta t$ from 1 to 12ms. We found that for the vast majority of the datasets, the median of the inter-spike interval (ISI) was $\Delta t=4$ms. Binning the data at $\Delta t=4$ms gave the  best power-law avalanche distributions in terms of scaling relation statistics.  

Avalanche degrees of freedom were determined in the standard way  (\cite{christensen2005,beggs2003neuronal}), i.e., by clusters of network activity bounded by times of no activity.
In the extracted clusters of activity, avalanche size $S$ is defined as the number of events within the cluster, and avalanche duration $T$ as the time interval between the first and last event in the cluster. After collecting all the avalanches for a given subject data, we calculated their probability distributions, $P(S)$ and $P(T)$, and their corresponding effective exponents $\tilde{\tau}_S$ and $\tilde{\tau}_T$. The distributions were logarithmically binned to reduce the noisy character of the distributions tails, as described in  \cite{christensen2005}, after which we fitted a line in log-log scale to find the exponents and its errors. After the exponents $\tilde{\tau}_S$ and $\tilde{\tau}_T$ were found, we plotted them on the $\gamma$-scaling line as shown in Fig. (\ref{fig:allData}).

We then used the so-called distance from criticality criterion developed in \cite{Ma2020} to keep data consistent statistically with quasicritical behavior. Essentially, this distance quantifies departure from a perfect scaling relation and was set to 0.2. 
We found that out of the 604 datasets, 566 satisfied this selection criterion.

\subsection{Effective critical exponents on the $\gamma$-scaling line}

We have seen in previous work \cite{fosque2021evidence} that the effective avalanche distribution exponents of a network with different parameters and noise will move along the $\gamma$-scaling line as long as the system lies in the quasicritical region. In this work, we
explore the potential relation between the position of exponents on the $\gamma$-scaling line and the age and gender of the subjects. We thus need a way to project the pair of effective exponents ($\tilde{\tau}_S, \tilde{\tau}_T$), defining a point, onto the $\gamma$-scaling line to quantitatively asses its position along the line. 

\begin{figure}[h!]
    \centering
    \includegraphics[scale=0.35]{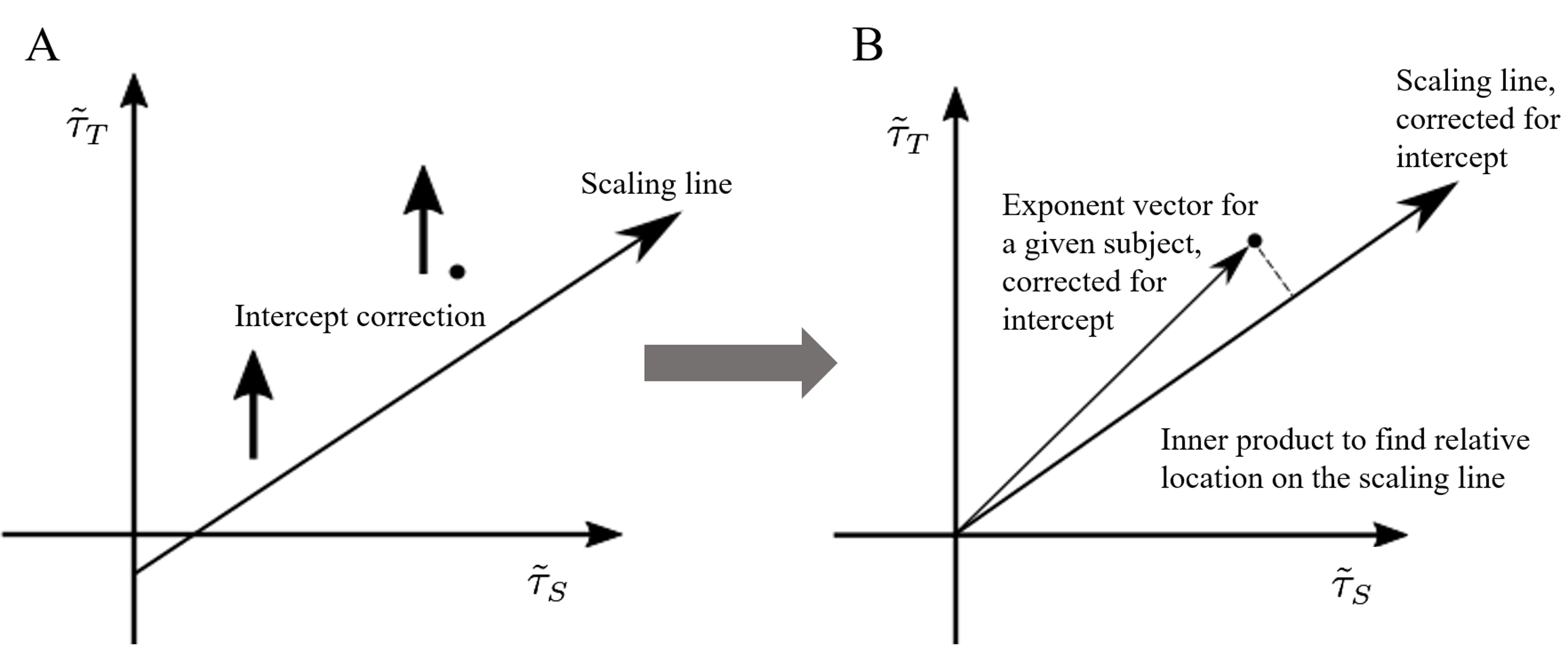}
    \caption{Pairs of size and duration effective exponents for each subject is transformed to a normalized value between 0 and 1. We call this measure the ``position on the $\gamma$-scaling line". This measure is obtained by \textbf{A:} shifting the $\gamma$-scaling line, and \textbf{B:} projecting 
    each subject's duration-size vector onto the shifted $\gamma$-scaling line. The largest projection represents the value of 1. } 
    \label{fig:posOnLine}
\end{figure}

To this end, once the $\gamma$-scaling line is determined, we consider the following 
projection method (see Fig. \ref{fig:posOnLine}):
\begin{itemize}
    \item Shift vertically all points along the $\tilde{\tau}_T$ axis, so that the $\gamma$-scaling line intercepts the origin $(0,0)$. Note that we are only interested in the relative position on the line, so shifting all points does not affect the results.
    \item Project the vectors, defined from the origin to the shifted points ($\tilde{\tau}_S, \tilde{\tau}_T$), onto the new $\gamma$-scaling line as indicated in Fig. \ref{fig:posOnLine} \textbf{B}. 
    \item Choose the projected point with largest distance from the origin on the shifted $\gamma$-scaling line. This distance will be defined as 1.
    \item Re-scale all other projected points by the maximal distance, so that all distances fall in the interval $[0,1]$.
\end{itemize}

 Consequently, how far from the origin a given subject's point lies on the $\gamma$-scaling line provides a measure whose magnitude may correlate to age and/or gender, thus providing a biomarker. 

\subsection{Determining branching ratio $\sigma$ and susceptibility $\chi$}

We used a recently-developed multi-step regression algorithm (MR. Estimator, \cite{priesemann2021}) to calculate the branching ratios, $\sigma$, of each subject. This algorithm allows to study subsampled time series and obtain the approximate, empirical, branching ratio $\sigma$. 

We also calculated the local time fluctuation (LTF) as described in \cite{fosque2021evidence}, which is similar to the coefficient of variation. The LTF measures the temporal fluctuations of the density of active nodes, or in a more colloquial sense, the variability of firing rate that the network displays at a given time interval. One starts by calculating the density of active nodes in the network at every time step, 
$$\rho_1(t)=\frac{1}{N}\sum_{i=1}^N \delta_{z_i(t),1},$$
where $N$ is the total number of nodes, $z_i(t)$ is the state of node $i$ at time $t$, and $\delta_{z_i(t),1}$ is 1 if $z_i(t)=1$ and zero otherwise. Then, we calculate the average firing rate for each subject by averaging the activity density over the total recording time $T$, 
$$\langle\rho_1(t)\rangle_T=\frac{1}{T}\sum_{t=1}^T \rho_1(t).$$
Notice that this definition of ``firing rate" can never be greater than one because we are considering the fraction of population active at time $t$. We can also calculate the (dynamical) susceptibility defined as 
\begin{align}
    \chi = N[\langle\rho_1^2(t)\rangle_T-\langle\rho_1(t)\rangle_T^2],
\end{align}
where $\langle\rho^2_1(t)\rangle_T=\frac{1}{T}\sum_{t=1}^T \rho^2_1(t)$.
Once the susceptibility, $\chi$, and average firing rate, $\langle\rho_1(t)\rangle_T$, have been obtained one can calculate the LTF,
\begin{align}
    {\rm LTF}=\frac{1}{\langle\rho_1(t)\rangle_T}\sqrt{\frac{\chi}{N}}.
\end{align}
As seen from this equation (3), if the standard deviation in average activity, $\sqrt{\chi}$, is large compared to 
the average firing rate, then ${\rm LTF}>1$, which translates into a bursting activity, whereas whenever ${\rm LTF}<1$ the activity becomes more regular.

\subsection{CBM: A minimal neural network model}

Computational models are often extremely helpful in interpreting data and building intuitions about different mechanisms in nature. In neuroscience different models are proposed to mimic a variety of dynamical processes involving neurons. We are interested in understanding the collective behavior of a large number of neurons. Here, the computational neuroscience community is divided into two major camps: rate-type and spike-type models. For statistical analysis of avalanches, discrete-time spike models are more appropriate. In addition, our neuronal network is interacting spatially and temporally, that is, the state of a neuron will depend on the state of its connected neighbors at a previous time. Since the transmission of information between connected neurons is intrinsically probabilistic, the model needs to be stochastic. These are known as probabilistic cellular automata models. 

Motivated by this, we employ a computational model based on our previous work (\cite{williams-garcia2014, fosque2021evidence}). Our cortical branching model (CBM), a probabilistic cellular automaton, makes use of the branching process to characterize avalanches and the spread of information across the network (\cite{williams-garcia2014}). The CBM is an elementary minimal model of cortex dynamics that encapsulates many of the experimentally relevant collective phenomena of neural networks. This model qualitatively captures the power law distributions of avalanches  seen experimentally in cortical slice networks (\cite{beggs2003neuronal}; \cite{beggs2005}).  We let each neural node have the same small probability $p_s$ of spontaneous activation, mimicking the effect of external noise. Note that this model can also be used to obtain critical dynamics by letting
$p_s=0$ and $\kappa=1$. The CBM is consistent with the organizing principle of quasicriticality but clearly many other models satisfy such principle.  

In this work, CBM simulations consisted of $N=256$ nodes, each having a number of incoming neighbors $k_{in}=5$, a connectivity matrix with weights $P_{ij}$, and a fixed refractory period $\tau_r=1$ during which the node is not able to  activate. Each node can be activated in two ways. First, it could be driven by its incoming neighbors. Second, it could activate spontaneously due to the external noise with probability $p_s=10^{-3}$. We will describe each of these next.

\underline{\it Driven activity}: Each connection from node $i$ to node $j$ has a weight $P_{ij}=\kappa p_{n_{ij}}$ indicating the probability of transmission of activation $p_{n_{ij}}$, $0\leq p_{n_{ij}}\leq 1$, randomly chosen from a weighting function (described below). The label $n_{ij}\in \{1, ..., k_{in}\}$ ranks each neighbor connection coming from neighbor $i$ to target node $j$ by strength, e.g., $n_{ij}=1$ corresponds to the strongest connection inbound at node $j$. The sum of probabilities emanating from each node $i$ adds to one:  
$$1=\sum_{i=1}^{k_{in}}  p_{n_{ij}}.$$

Note that, when the network is tuned to be critical, the sum of incoming probabilities is the branching ratio $\sigma$ by construction.  We incorporate the branching parameter $\kappa$ to modulate the state of the network. The relation between the branching ratio $\sigma$ and the branching parameter $\kappa$ is explained below. Node $j$ at time step $t+1$ becomes active if node $i$ in the previous time step $t$ was active and the connection between them transmitted. A connection from $i$ to $j$ transmits if ${\rm rand} \leq P_{ij}$,  where $\rm rand$ is a uniformly distributed random number drawn from the interval [0, 1].  

Processing nodes update at each time step to simulate the propagation of activity through the network. After becoming active, a node would become inactive, or refractory, for the next $\tau_r=1$ time steps. 

\underline{\it Branching}: The branching parameter $\kappa$ plays the role of the control parameter in the CBM, similar to  temperature in the Ising model. This parameter places the system in or out of the quasicritical region. In the critical case, when $p_s=0$, $\kappa=1$ would put the system in the critical region. However, in the quasicritical case $\kappa$ will have different values  depending on the external noise $p_s$ and other parameters of the network. 

\underline{\it Spontaneous activity}: In addition to driven activity from neighbors, each node has a small probability, given by $p_s$, of becoming spontaneously active at any time step.  The network can only be critical when $p_s$ = 0. As $p_s$ increases, the network activity moves into the quasicritical region.  

\underline{\it Weighting function}: As many studies report weight distributions with a nearly exponential form [\cite{chen2010}, \cite{brunel2004},\cite{barbour2007}] , we adopt a weighting function where transmission probabilities $p_{n_{ij}}$ are determined by the following equation
$$p_{n_{ij}} = \frac{e^{-Bn_{ij}}} {\sum_{n=1}^{k_{in}} e^{-Bn}},$$
where $i$'s are neighbors to the target $j$ node. 
When the bias exponent $B\ge0$ equals zero, the distribution is homogeneous; as $B$  increases, the distribution of $p_{n_{ij}}$ values becomes increasingly inhomogeneous with few strong connections and many weak ones. Because data suggests that connection strengths become increasingly skewed with age, we wanted to be able to capture this in the model.  

We increased $B$ to simulate the changes in connections reported in ageing brains. Our justification for this was motivated by the results of (\cite{otte2015aging}), who looked at connection strengths in human patients as they aged. There, the density of fiber bundles was assessed by Diffusion Weighted Imaging (DWI). The entire cortical mantle was parcellated into regions (nodes) and the strength of each node was taken as the sum of connection strengths into and out of a region. When average node strengths were arranged in descending order and then separated by age groups, it revealed that curves from older patients dropped more steeply than for younger patients. The best fit exponential curves for the three age groups revealed a trend, where the exponents became larger for older cohorts. For ages 25-45: 0.02338  (0.02043, 0.02633), ages 45-65: 0.02909  (0.0257, 0.03248), and ages 65-90: 0.03067  (0.02601, 0.03534); figures given as mean value, followed by $95\%$ confidence limits. 
\begin{figure}[h!]
    \centering
    \includegraphics[width=0.5\linewidth]{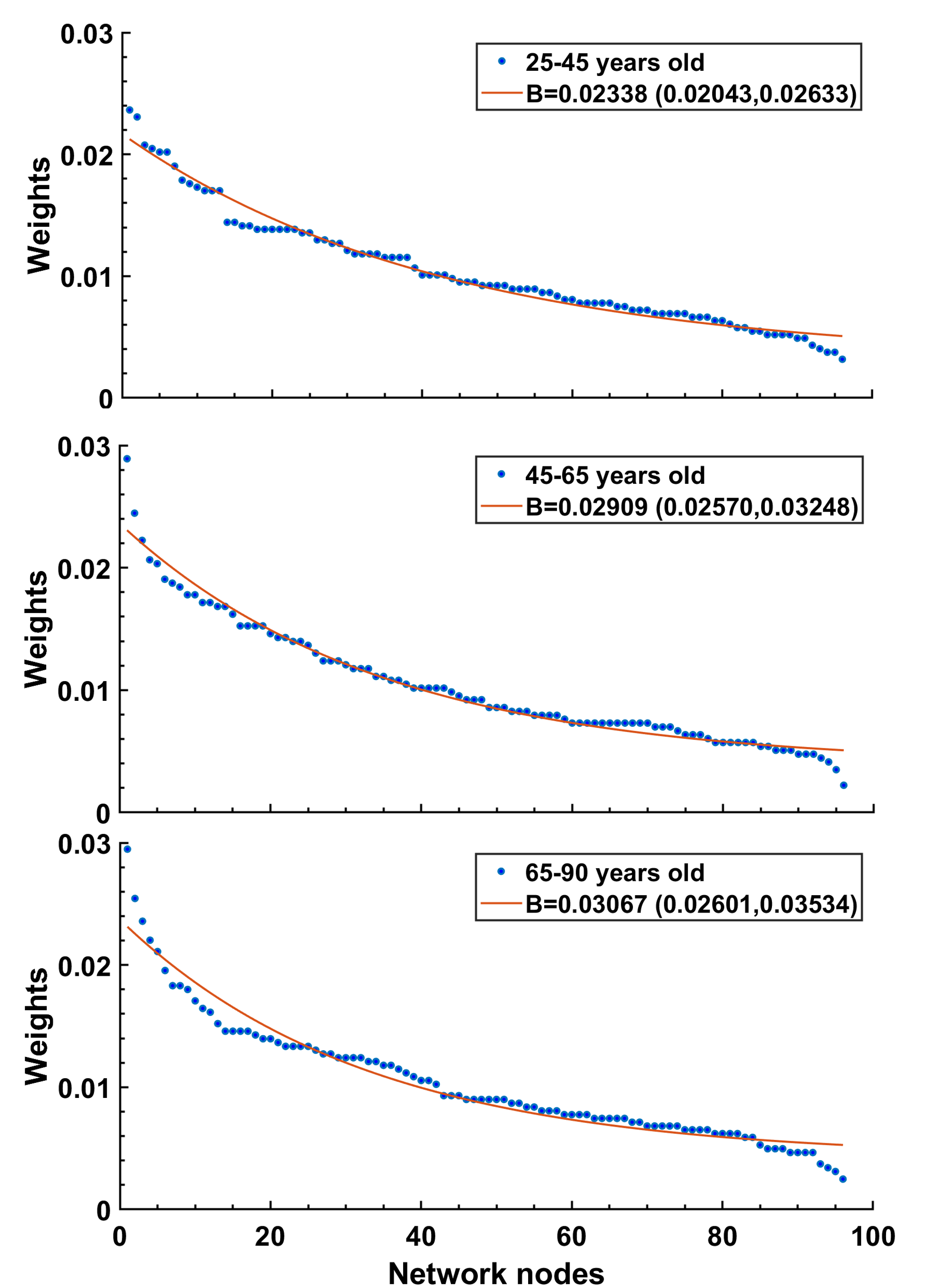}
    \caption{Human data suggests that exponential decay of connection strengths becomes steeper with age. Figure adapted from (\cite{otte2015aging}). Note that oldest age group (bottom) shows the sharpest drop, while the youngest age group (top) has the shallowest curve. For details, see text.}
    \label{fig:weightsOtte}
\end{figure}

\section{Results}

\subsection{$\gamma$-Scaling line results}

\begin{figure}[h!]
    \centering
    \includegraphics[scale=.4]{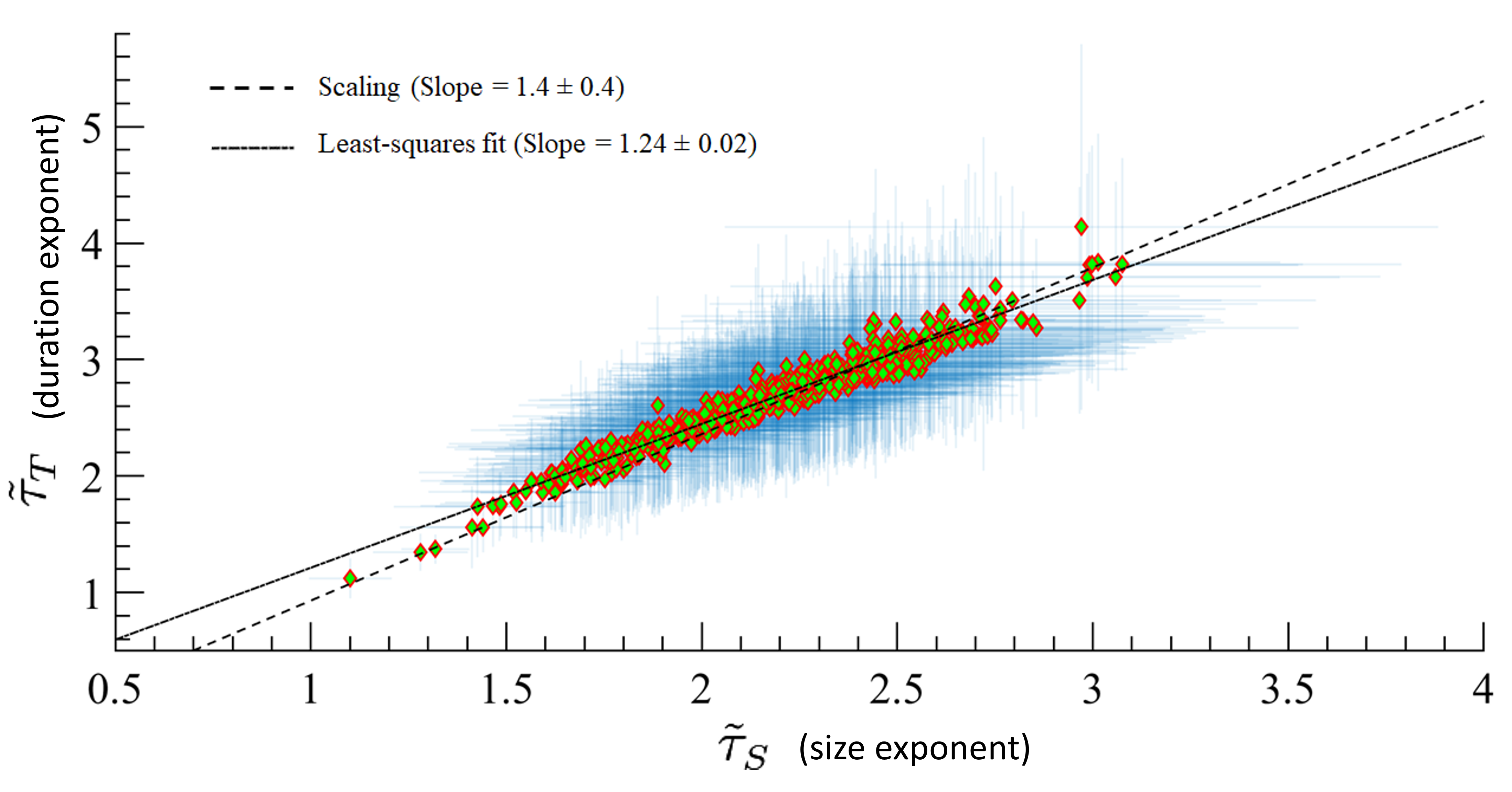}
    \caption{MEG data from 566 human subjects on the $\gamma$-scaling line with ages ranging from 18 to 88 years old. Error bars are shown in blue. A least-squares fit of exponent pairs is also shown in dashed line, and gives a least-squares fit of $\gamma_{\text{lsf}}=1.24\pm 0.02$. Solid line indicates the average scaling slope,  $\langle\frac{\tilde{\tau}_T-1}{\tilde{\tau}_S-1}\rangle =1.4\pm 0.4$. The scaling slope is the average of scaling fraction across subjects.}
    \label{fig:allData}
\end{figure}

A major prediction of quasicriticality is that human MEG data should organize along a $\gamma$-scaling line just as has been seen previously with animal's spiking data (\cite{fontenele2019criticality}; \cite{fosque2021evidence}). To test this, we calculated effective exponents from the avalanche distributions of the included MEG data and found, as predicted, that all these exponents are distributed along the $\gamma$-scaling line.  The average $\gamma$ exponent for all datasets is $\langle\gamma\rangle=1.07 \pm 0.03$, and the average scaling $\langle\frac{\tilde{\tau}_T-1}{\tilde{\tau}_S-1}\rangle =1.4 \pm 0.4$. Recall that $\gamma$ is obtained from the power-law relation between the average avalanche size distribution $\langle S \rangle$ and the avalanche duration $T$. These results are consistent within statistical error bars, and we need to remember that the scaling fraction is sensitively dependent upon errors of the measured effective exponents, so larger errors are to be expected.  Figure (\ref{fig:allData}) shows all datasets on the $\gamma$-scaling line with their corresponding error bars. These results show a similar characteristic distribution of effective exponents as in \cite{fontenele2019criticality} and \cite{fosque2021evidence} although with a slightly different slope. Hence, we see that these results appear to be consistent with the organizing principle of quasicriticality.

\begin{figure}[h!]
\centering
\includegraphics[scale=0.5]{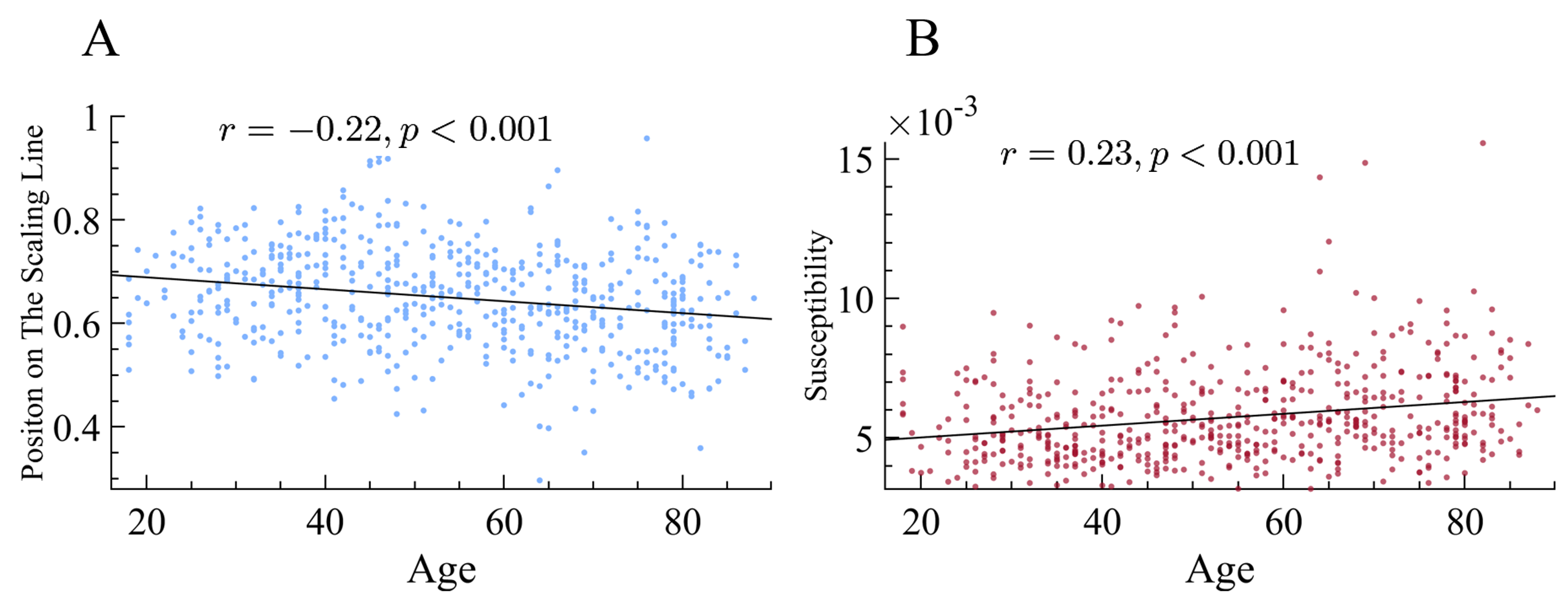}
\caption{Older subjects display smaller exponents and higher susceptibility. \textbf{A:} Older subjects have exponents that fall lower on the $\gamma$-scaling line. We found that there is a significant negative correlation between age and the magnitude of time and size exponents in our dataset. \textbf{B:} Older subjects have higher dynamical susceptibility. There is a significant correlation between age and susceptibility. }
\label{fig:f4}
\end{figure}

After confirming that the data followed the principle of quasicriticality, we next sought to find out if the position on the $\gamma$-scaling line could reveal information about the patient population. We first looked at the differences between age groups with respect to their exponents. We used the distance measure explained earlier and found that there is a small but significant negative correlation between age and the position on the line. More specifically, being older is found to be negatively correlated with the magnitude of the duration and size exponents (Fig. \ref{fig:f4}) and therefore, a lower location on our previously defined position on the $\gamma$-scaling line. 


\begin{figure}[h!]
\centering
\includegraphics[scale=0.35]{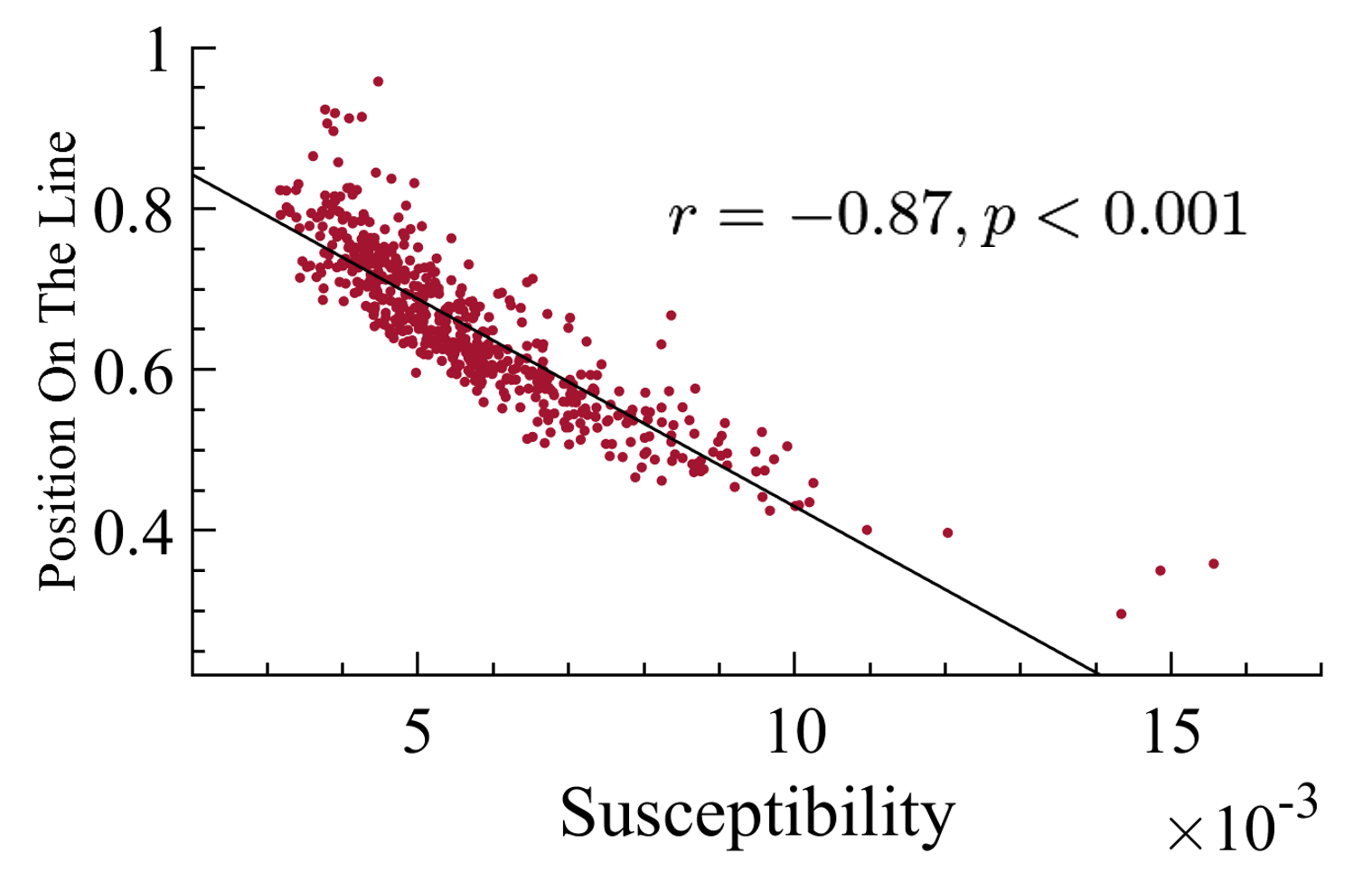}
\caption{There is a negative correlation between position on the line and susceptibility. This correlation suggests that subjects with smaller exponents have higher susceptibility (correlation slope $r=-0.87$, $p < 0.001)$.}
\label{fig:chiPosOnline}
\end{figure}

Another prediction of quasicriticality is that when $p_s$ increases, susceptibility decreases (as long as the system is near the peak of maximum susceptibility), and exponents get smaller, which translates into a position closer to the origin on the $\gamma$-scaling line. Hence, after finding the differences in position on the $\gamma$-scaling line with respect to age, and the relation with their exponents, we decided to analyze their susceptibility. We found that there is a strong and negative correlation between susceptibility and position on the line. In other words, subjects with smaller effective exponents show higher susceptibility in their MEG data, see Fig. (\ref{fig:chiPosOnline}) for details. Since older subjects have lower position on the line, this means that the older subjects also have higher dynamical susceptibility, see Fig. (\ref{fig:f4}). This result may seem to contradict the prediction of quasicriticality, but recall that the susceptibility depends on multiple parameters of the network. We will show below a potential parameter that may be causing this trend on the $\gamma$-scaling line.


\begin{figure}[h!]
\centering
\includegraphics[scale=0.4]{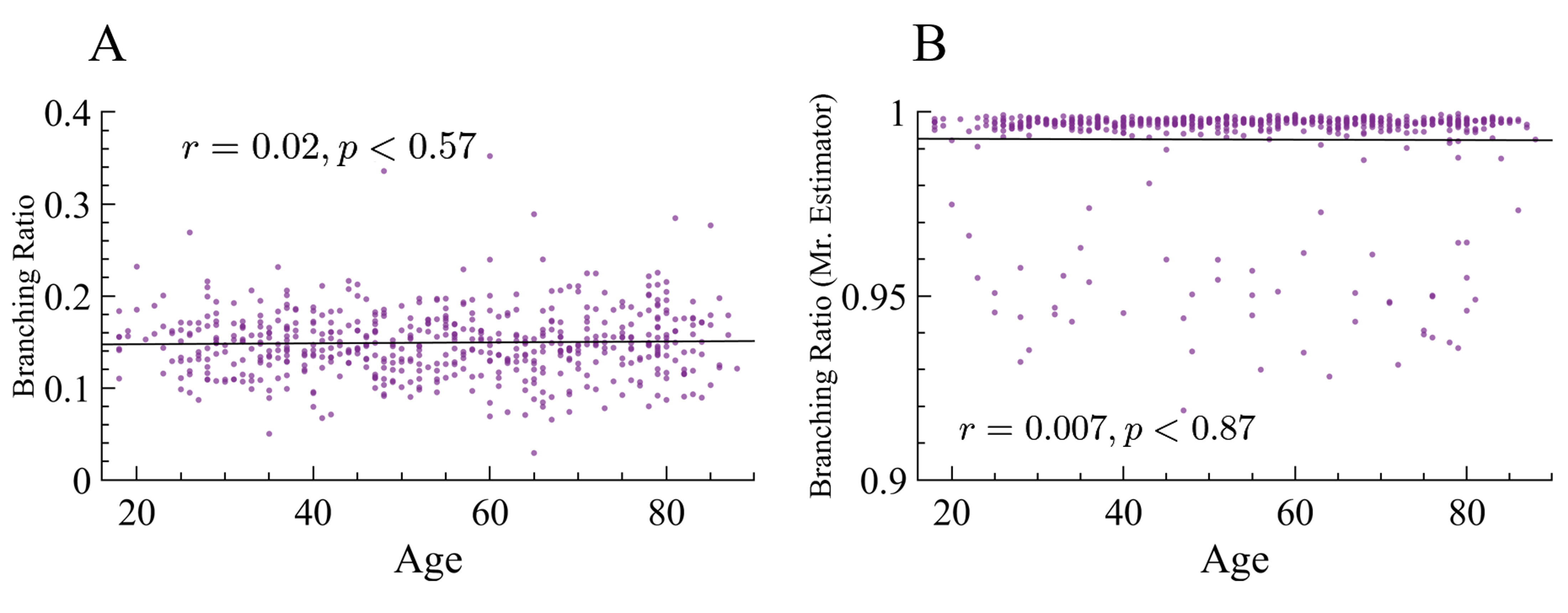}
\caption{Branching ratio does not significantly change with age. \textbf{A:} Naive method: $r= 0.02$, $p< 0.57$. \textbf{B:} MR. Estimator : $r= - 0.007$,
$p< 0.87$.}
\label{fig:branchingRatioFigs}
\end{figure}

As stated in the Background section, there is an important relation between susceptibility and branching parameter $\kappa$. This relation states that the peak of maximum susceptibility will move towards lower values of the branching parameter as $p_s$ is increased. Our analysis of susceptibility showed that older subjects have higher values of susceptibility than younger ones. Since we cannot extract the parameter $\kappa$ from this data, and we know that there is a relation between $\kappa$ and $\sigma$, we determine $\sigma$ instead experimentally.
 However, one of the most striking results is the fact that the branching ratios did not significantly change across age groups. We used both MR. Estimator and naive methods and find that, regardless, this parameter does not change with age, see Fig. (\ref{fig:branchingRatioFigs}) for more details.

\begin{figure}[!h]
\centering
\includegraphics[scale=0.4]{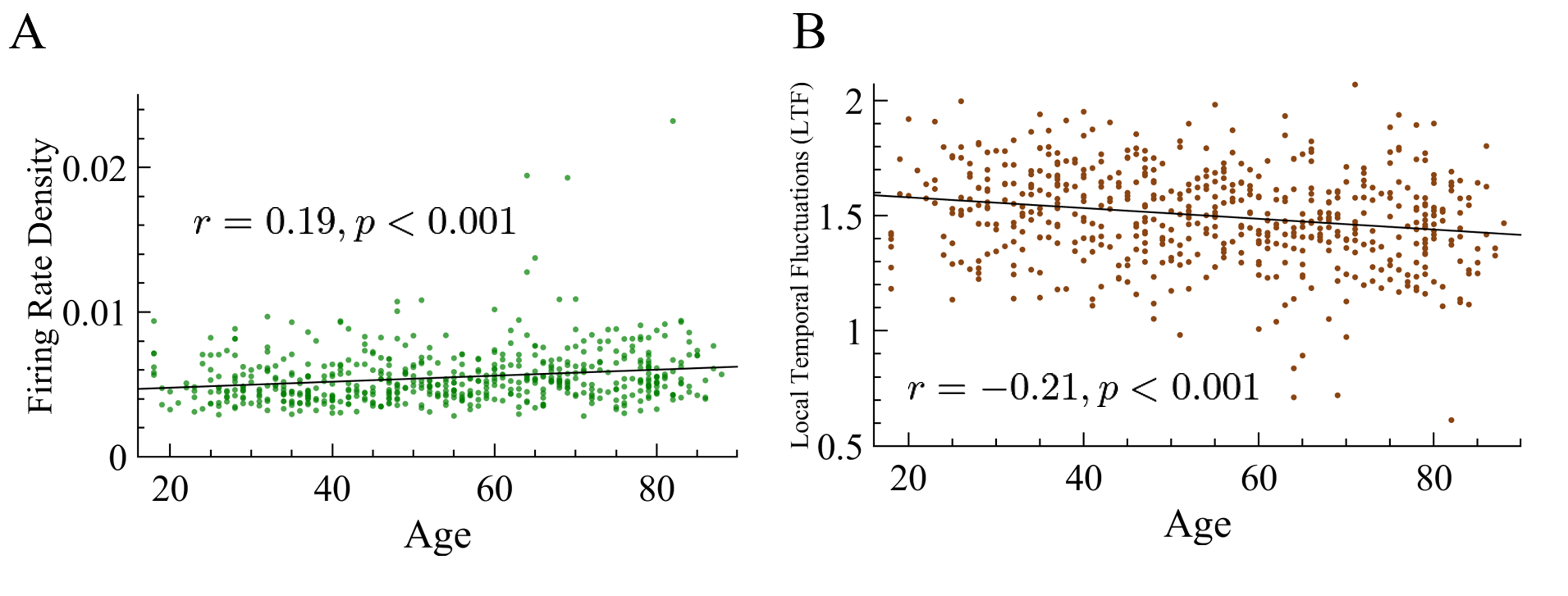}
\caption{Age positively correlates with firing rate density and shows a negative correlation with LTF. \textbf{A:} Firing rate density vs age shows a significant positive correlation ($r=0.19$, $p<0.001$). The statistical significance of this relationship is not dependent on the three outlier values that are above 0.015. \textbf{B:} LTF vs age shows a significant negative correlation ($r=-0.21$, $p<0.001$).}
\label{fig:f5}
\end{figure}

Quasicriticality also predicts that with increasing external noise there will be an increase of avalanche concatenation, which in turn would result in smaller exponents. We mentioned before that the LTF, as the coefficient of variation, measures the level of variability in activity in the network. Similarly, the firing rate of a network can give clues on the level of activity in the subject given that an increase in the firing rate increases the probability of larger avalanches. Hence, we also analyzed the variability in activity across the different age groups by calculating the LTF and the firing rate density for each subject as defined in Methods. LTF is found to be reduced with age, and the firing rate density is found to increase with age. Although we see variability around the average, the trend is very consistent, $p<0.001$ in both cases, see Fig. (\ref{fig:f5}) for more details. It is important to remember that an increase in external noise will result in a lower value of LTF, but a low value of LTF does not necessarily implies an increase in $p_s$ since there are other parameters that can influence this variability.

\begin{figure}[h!]
    \centering
    \includegraphics[scale=0.25]{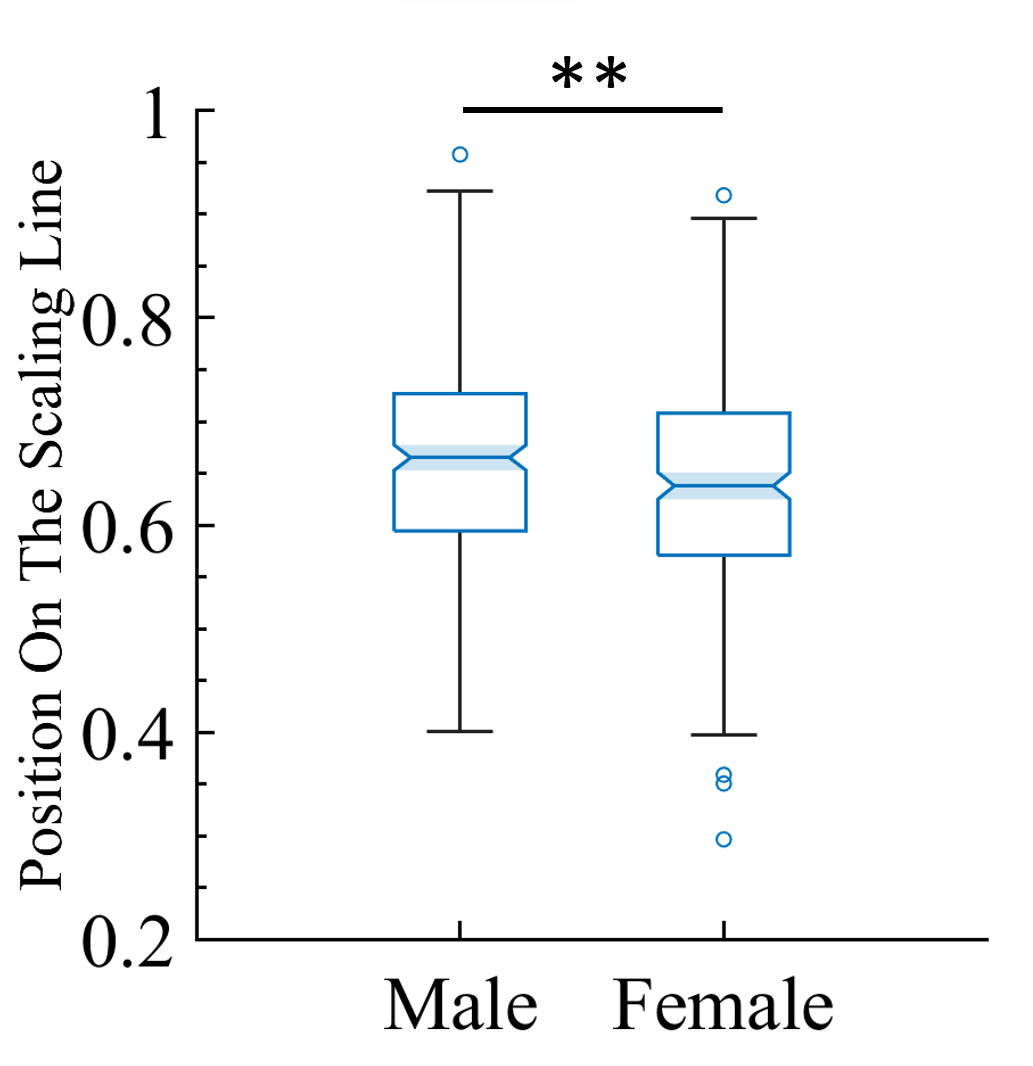}
    \caption{There is a small but significant difference in position on the line between male and female subjects. The box plot shows the distribution of the female and male subjects' data and where they land on the line. For female subjects (mean=0.64, SD=0.10), while for male subjects (mean=0.66, SD=0.09). The two-tailed t-test gives ${\sf t}=-2.97$ and $p<0.01$ (indicated by the two stars).}
    \label{fig:gender2}
\end{figure}

One of the goals of this study is to find potential biomarkers by using the framework of quasicriticality. Thus, given the vast amount of information contained in this large MEG dataset, we decided to see if we could use our framework to predict gender from the data. We used the position on the $\gamma$-scaling line measure to investigate whether the magnitude of the duration and size exponents are different between male and female subjects. Indeed, there is a significant difference, suggesting that this approach may hold promise as a biomarker.  We found that male subjects have a slightly higher position magnitude compared to female subjects, see figure (\ref{fig:gender2}). 


\subsection{CBM Model fits the age related reduction in exponents}

\begin{figure}[h!]
    \centering
    \includegraphics[width=0.8\linewidth]{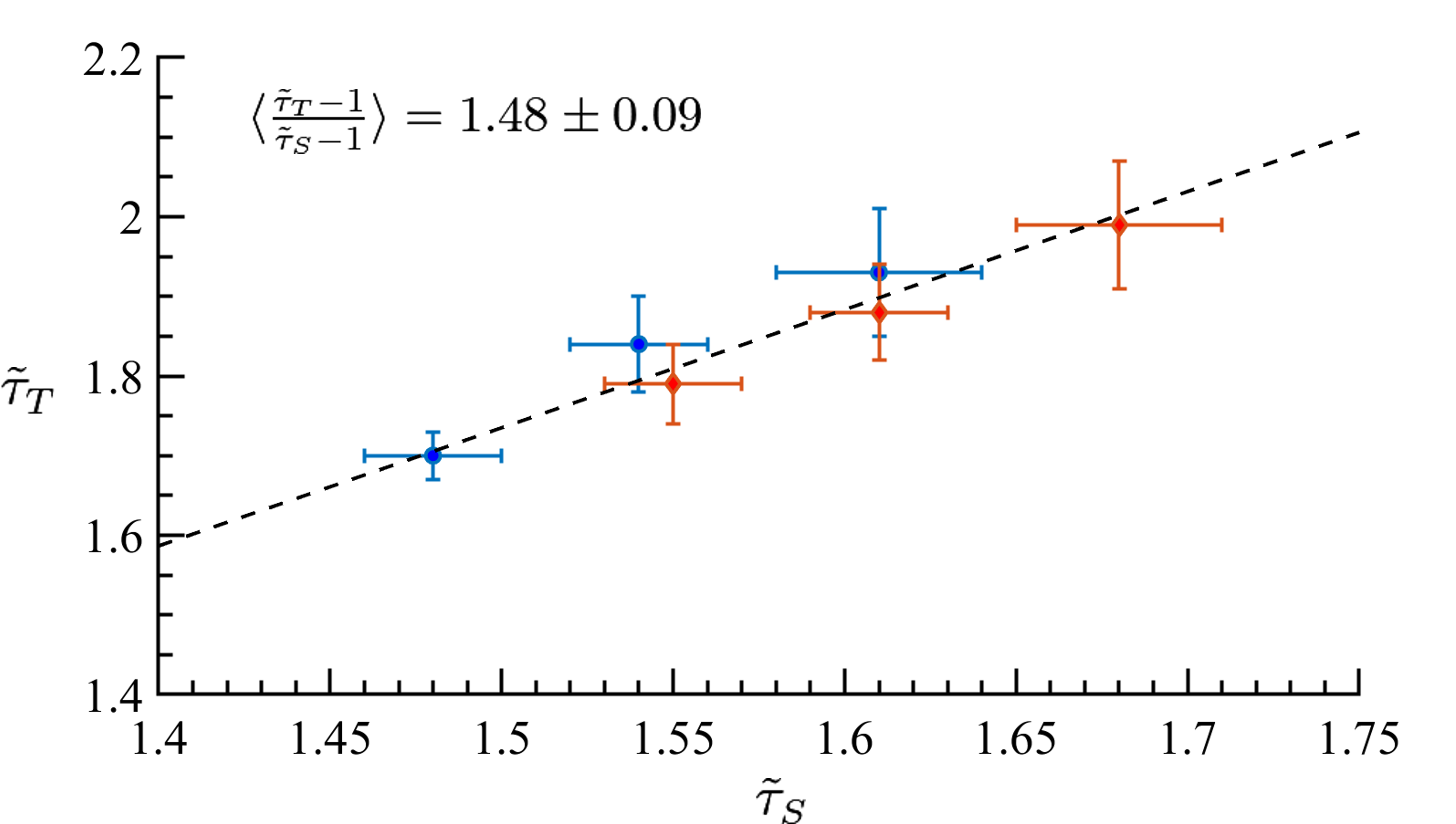}
    \caption{Simulation on the $\gamma$-scaling line. Probability of spontaneous activation $p_s=10^{-3}$, and bias parameter $B=0.6$ (red diamonds) and $B=1.8$ (blue circles). The three points for each bias represent three different $\kappa$ values around the peak of maximum susceptibility taken from tables \ref{table1} and \ref{table2}. We can see that as we increase the bias parameter, the exponents get smaller while increasing susceptibility.}
    \label{fig:simScaling}
\end{figure}

Motivated by the lower exponent values among older subjects, we sought to create a model that can account for this experimental finding. As mentioned in Methods previous literature \cite{otte2015aging} indicates that natural/healthy aging involves the weakening of weak weights and enhancement of strong connections in functional connectivity. We ran simulations with our CBM with different weight distributions to mimic this trend. In our CBM the steepness of these distributions is characterized by the bias parameter $B$. We ran CBM simulations with bias parameter $B=0.6$ and $B=1.8$. Recall from Methods that a higher value of $B$ means a more positively skewed distribution of weights while a lower value of $B$ would correspond to a flatter distribution. We observed that simulations with a bias of $B=0.6$ have a lower susceptibility peak but larger exponent values than the simulations with larger bias. These simulations were run under the same probability of spontaneous activations, $p_s=10^{-3}$, and on a network with 256 nodes, 5 incoming neighbors, and refractory period of one time step. These results indicate that making connectivity weights steeper reduces the size and duration exponents (Fig. (\ref{fig:simScaling}) and tables \ref{table1} and \ref{table2}). These results from the CBM simulations fit the trend of the MEG experimental results where there is an overlap of exponents along the $\gamma$-scaling line whenever the system is around the peak of maximum susceptibility.

\begin{table}[h!]
\centering
\begin{tabular}{|r|r|r|r|r|r|}
\hline
\multicolumn{1}{|l|}{$\kappa$} & \multicolumn{1}{l|}{$\chi$} & \multicolumn{1}{l|}{$\tilde{\tau}_S$} & \multicolumn{1}{l|}{$\tilde{\tau}_T$} & \multicolumn{1}{l|}{$\frac{\tilde{\tau}_T-1}{\tilde{\tau}_S-1}$} & \multicolumn{1}{l|}{$\gamma$} \\ \hline
1.06 & 0.9 $\pm$ 0.2 & 1.48 $\pm$ 0.02 & 1.70 $\pm$ 0.03 & 1.46$\pm$ 0.03 & 1.52$\pm$ 0.02 \\ \hline
1.07 & 0.9 $\pm$ 0.1 & 1.54$\pm$ 0.02 & 1.84$\pm$ 0.06 & 1.56$\pm$ 0.05 & 1.54$\pm$ 0.02 \\ \hline
1.08 & 0.9$\pm$0.2 & 1.61$\pm$ 0.03 & 1.93 $\pm$ 0.08 & 1.54$\pm$ 0.07 & 1.52$\pm$ 0.02 \\ \hline
\end{tabular}
\caption{Table of exponents from CBM simulations for bias $B=1.8$ for $\kappa$ values around the peak of maximum susceptibility.}
\label{table1}
\end{table}

\begin{table}[h!]
\centering
\begin{tabular}{|r|r|r|r|r|r|}
\hline
\multicolumn{1}{|l|}{$\kappa$} & \multicolumn{1}{l|}{$\chi$} & \multicolumn{1}{l|}{$\tilde{\tau}_S$} & \multicolumn{1}{l|}{$\tilde{\tau}_T$} & \multicolumn{1}{l|}{$\frac{\tilde{\tau}_T-1}{\tilde{\tau}_S-1}$} & \multicolumn{1}{l|}{$\gamma$} \\ \hline
1.11 & 0.44 $\pm$ 0.02 & 1.55$\pm$ 0.02 & 1.79$\pm$ 0.05 & 1.44$\pm$ 0.04 & 1.53 $\pm$ 0.02 \\ \hline
1.12 & 0.44$\pm$ 0.01 & 1.61$\pm$ 0.02 & 1.88$\pm$ 0.06 & 1.44 $\pm$0.05 & 1.53 $\pm$ 0.02 \\ \hline
1.13 & 0.43$\pm$0.02 & 1.68$\pm$ 0.03 & 1.99$\pm$ 0.08 & 1.46$\pm$ 0.06 & 1.52$\pm$0.02 \\ \hline
\end{tabular}
\caption{Table of exponents from CBM simulations for bias $B=0.6$ for $\kappa$ values around the peak of maximum susceptibility.}
\label{table2}
\end{table}

Thus, our computational model can capture the trends observed in the data when it is supplied with similar connectivity parameters. 

\section{Discussion}

We began this paper by suggesting that neural biomarkers could be based on quantities that are homeostatically regulated for health, like functional neuronal activity. The principle of quasicriticality predicts that neural networks will operate in a quasicritical region associated to a critical point, even in the face of perturbations, by moving along a dynamical scaling line. We hypothesized that a patient’s position along this scaling line could therefore contain valuable health information about their relation to their ``normal state’’ and various factors that might perturb them from it. Here we used a large MEG data set to conduct a first test to see if this quasicritical framework could show potential for developing biomarkers. Our first main finding was that the majority of MEG data indeed fell along the scaling line. The second main finding was that the position on this line was significantly related to the age of the subject and their gender. The third main finding was that the susceptibility of the subjects, akin to their sensitivity to new stimuli, was also significantly related to their position along the scaling line. To make sense of these findings, we employed a previously published network model, the cortical branching model. When the model was supplied with connectivity data that reflected the age of the patients, it qualitatively reproduced the results found in the data – this is our fourth main finding. Quasicriticality can thus offer a plausible explanation as to why older subjects become more easily distracted (higher susceptibility) as their connections change with age. We suggest that the quasicritical framework therefore should be further investigated as a platform for developing biomarkers of neurological health.  

Previous work on human MEG data showed that it contained signatures of criticality in the form of avalanche distributions that followed power laws, both for spontaneous activity (\cite{shriki2013neuronal}) and stimulus-evoked activity (\cite{Arviv2015}). The work by Arviv and colleagues even examined the scaling relation but did not find it to be well satisfied. They speculated that this was because there were not enough stimulus-evoked data to conclusively evaluate its validity (\cite{Arviv2015}). In the present work, we expanded on these pioneering results by exploring an MEG data set of spontaneous activity that was almost thirty times larger ($n = 21$ vs $n = 604$) and, at the same time, we broaden the framework to that of quasicriticality, a more explanatory framework. Because of this, we were able to examine the scaling relation more fully, finding that it fit within experimental error for the vast majority of patients. 

Many other groups have noted connections between criticality and neurological health. Broadly speaking, these works have used distance from the critical point, variously assessed by the branching ratio, the quality of the power laws/avalanche shape collapse, the extent of multifractality, or the degree to which the exponent relation is satisfied as the relevant variables. For example, this approach has been taken with respect to sleep and sleep deprivation (\cite{Meisel2013,Priesemann2014}), epilepsy (\cite{Meisel2012,Meisel2015,Arviv2016,Hagemann2021}), hypoxia (\cite{Roberts2014}), stroke (\cite{Rocha2022}), schizophrenia (\cite{Alamian2022}) and Alzheimer’s disease (\cite{Jiang2018}), to name a few. For overviews, see (\cite{Massobrio2015, Zimmern2020,Fekete2021}).

Although our present work shows a few similarities with these previous studies, we are the first, to our knowledge, to indicate how the position along the scaling line may be used as a potential biomarker. Strictly speaking, our work does not evaluate distance to the critical point, but rather how the system configures itself in exponent space as it continues to maintain its activity homeostatically regulated to remain in the quasicritical region close to the Widom line, i.e., the line of maximal susceptibility. More concretely, two patients could be equally close to the critical point by some measure but could still lie on different portions of the scaling line. This illustrates that we are taking a qualitatively different perspective from that of previous studies, one that could potentially reveal new information. 

Because this work was only a first step toward using the principle of quasicriticality to develop biomarkers, it has room for improvements. First, the MEG recordings were rather brief, averaging about 9 minutes per subject, which curtailed the statistical power of our analyses. In the future, it would be better to have longer recordings (30 – 60 mins) with more neuronal avalanches. Second, the MEG scanner used to produce this data set provided 102 time series. While this is state of the art, in the future we would like to increase the number of channels to enhance statistics. Third, in this work we used a data set that had only the most basic biological information like age and gender; this restricted the types of conclusions that we could draw. A data set that contained information about multiple health disorders like the degree of dementia, epilepsy, depression or schizotypy would have allowed many dimensions of neurological health to be checked for relationships with the position of each subject in exponent space along the scaling line. These proposed improvements should be addressed in future work. 

Very generally, the concept of criticality implies that a system must have a parameter that is precisely tuned to the critical point. In contrast, quasicriticality allows multiple parameters to interlock in specific ways to confine the system to a line or surface that could be considered approximately critical. Consistent with this, several studies have shown that as neural systems remain close to criticality, their avalanche distribution exponents vary in a peculiar way (\cite{shew2015turtle,fontenele2019criticality,Ma2020}). Specifically, the exponent $\gamma$ (for average size against duration) typically remains nearly fixed, while $\tilde{\tau}_S$ (for size) and $\tilde{\tau}_T$ (for duration) may vary but continue to nearly satisfy the exponent relation. The system is thus confined to move along a line. The fundamental reason behind a nearly constant $\gamma$ and why $\tilde{\tau}_S$ and $\tilde{\tau}_T$ are the exponents that vary is not known; these are intriguing open questions.

But because of these facts, we are now able to explore how nearly critical combinations of $\tilde{\tau}_S$ and $\tilde{\tau}_T$ relate to biological features. In the present work, we made first steps by linking them to age, gender and susceptibility. Now that this has been established, future work could investigate how these exponents relate to features of neurological disorders, as well as health features like resilience, creativity or intelligence.

To illustrate the potential utility of this, let us consider two scenarios: (1) departures from normality, and (2) dynamics of recovery. For (1), individual patients may have a tendency to spend more time on a specific region of the scaling line. Physicians could track this yearly as an indicator of what is typical for that patient. Sudden departures from that region of the scaling line could then signal that the patient’s brain is trying to compensate for some new neurological stress. The presence of this stress could be revealed in a yearly checkup by measuring the patient’s exponents. For (2), just as a cardiologist may apply a difficult treadmill challenge to a patient to stress test their heart and watch its recovery, a neurologist could apply a cognitive challenge to a patient and watch how the exponents move along the scaling line to assess recovery. The dynamics of how the exponents move back toward their original position may reveal the strength of homeostasis in that patient.

\section*{Conflict of Interest Statement}

The authors declare that the research was conducted in the absence of any commercial or financial relationships that could be construed as a potential conflict of interest.

\section*{Author Contributions}

L.J.F. performed the simulations and carried out the analysis of data and simulations. J.M.B., L.J.F., and A.A. presented the idea of connectivity and aging. M.Z. contributed with the MEG data and started the study. R.V.W.-G. contributed with the CBM code. G.O. developed the theoretical formalism in relationship to quasicriticality. G.O. and J.M.B. developed the concepts and wrote the manuscript. All authors discussed the results of the analysis and read, commented, and contributed to the content of the paper.


\section*{Funding}

Marzieh Zare was supported through funding from - IVADO postdoctoral fellowship program - PD-2019a-FQ-10, Montreal, CA. John M. Beggs was supported through NSF grant 2123781 (subcontract to JMB).  


\section*{Acknowledgments}
The authors would like to thank Naruepon Weerawongphrom for confirming the CBM simulations.


\section*{Data Availability Statement}
The datasets analyzed for this study can be found in the CamCAN repository \href{https://camcan-archive.mrc-cbu.cam.ac.uk/dataaccess/}{}.

\bibliographystyle{frontiersinSCNS_ENG_HUMS} 
\bibliography{test}

\end{document}